\documentclass[%
 reprint,
superscriptaddress,
amsmath,amssymb,
aip,
]{revtex4-2}

\usepackage{graphicx}
\usepackage{dcolumn}
\usepackage{bm}
\usepackage{ulem} 
\usepackage{orcidlink}
\usepackage{hyperref}

\newcommand{\iitj}{\affiliation{Department of Physics, Indian Institute of Technology Jodhpur\\
		N.H. 62, Nagaur Road, Karwar, Jodhpur, Rajasthan, India - 342030.}}

\newcommand{\iiserk}{\affiliation{Department of Physical Sciences, Indian Institute of Science Education and Research Kolkata\\
Mohanpur, West Bengal 741246, India}}
\begin{document}

\preprint{ }

\title{Tunable Interfacial Thermal Conductance in Graphene/Germanene van der Waals Heterostructure using an Optimized Interlayer Potential}

\author{Sapta Sindhu Paul Chowdhury$\,$\orcidlink{0009-0007-0472-7660}}
\email{saptasindhu62@gmail.com}
\iitj
\author{Sourav Thapliyal$\,$\orcidlink{0009-0003-3214-9184}}
\iitj
\author{Bheema Lingam Chittari$\,$\orcidlink{0000-0002-5868-2775}}
\iiserk
\author{Santosh Mogurampelly$\,$\orcidlink{0000-0002-3145-4377}}
\email{santosh@iitj.ac.in}
\iitj



\date{\today}

\begin{abstract}

Accurately modeling interfacial thermal transport in van der Waals heterostructures is challenging due to the limited availability of interlayer interaction potentials. We develop a pairwise interlayer potential for graphene/germanene van der Waals heterostructure using the binding energy obtained from \textit{ab-initio} density functional theory calculations and use it to calculate the interfacial thermal conductivity. Our calculations reveal that the interfacial thermal conductivity shows superior tunability with external strain. The phonon density of states calculations show a blueshift in the phonon spectra with an applied compressive strain in the direction of heat flow, increasing the interfacial thermal conductance to $\sim$136\% of the unstrained value. In contrast, a tensile strain is found to cause an opposite effect, reducing the conductance to $\sim$70\% of the unstrained value. Moreover, due to increased availability of phonons for heat transfer, both temperature and interaction strength are found to correlate positively with the interfacial thermal conductance for both directions of heat flow.
\end{abstract}

\maketitle

\section{\label{sec:intro}Introduction\protect} 
Understanding interfacial thermal conductance (ITC) is critical for advancing nanoscale heat management in emerging two-dimensional (2D) material-based technologies \cite{Chen2022}. As devices continue to scale down in size, the role of interfaces in dictating overall thermal behavior becomes increasingly dominant. 2D heterostructures exhibit interfacial heat transport behavior that deviates significantly from bulk systems, often dominated by weak van der Waals interactions and anisotropic phonon spectra \cite{Ong2012}. The study of ITC in two-dimensional materials has gained significant attention because of their promising applications in electronics, thermoelectric devices, and energy storage systems \cite{Zhang2018, Zong2024}.

Graphene is one of the most studied materials due to its superior physical properties \cite{Novoselov2012, Geim2007}. Monolayer graphene possesses one of the highest thermal conductivity of 2000-5000 W/m.K at room temperature \cite{Balandin2008}. As a result, graphene is considered to be a primary candidate for thermal interfacing material. Enormous effort has been made to understand the performance of graphene as a thermal interfacing material \cite{Zhang2018}. Ding et al. studied the ITC in graphene/MoS$_2$ heterostructure and found that due to the low friction at the graphene/MoS$_2$ interface, the thermal conductance is much lower than that of graphene/graphene and MoS$_2$/MoS$_2$ interfaces \cite{DING2016}. Unlike the lattice thermal conductivity, ITC is found to increase with temperature \cite{Liu2022}. The graphene/diamine heterostructure is investigated for ITC with variations of different physical parameters such as temperature, strain, stacking pattern, etc. It was found that the tensile strain negatively impacts the ITC \cite{Hong2022}. Among group IV materials, silicene-based interfaces with graphene were studied extensively. It was found that the ITC is higher when the heat conduction direction is from graphene to silicene compared to the reverse flow of heat \cite{Liu2014}. The strength of the van der Waals (vdW) interaction also impacts the ITC in this heterostructure \cite{zhou2022phonon}. Moreover, the ITC of silicene and different substrates, such as crystalline and amorphous silicon and silica, is found to be affected by temperature, crystalline nature of substrate, and strength of interaction \cite{Zhang2015itc}.

Although current mature electronics technologies are dominated by silicon, germanium is used in many of the essential applications in high-speed electronics, solar cells, and quantum technologies \cite{Hiller2013, Rodriguez2024, Mehmood_2024, Shi2021}. Germanene is the 2D counterpart of germanium and is fundamentally different from flat 2D materials like graphene \cite{expt-str}. Due to its low-buckled structure with heavier atomic mass, germanene shows distinctive thermal conductive behavior. The unique out-of-plane buckled structure of germanene amplifies these effects by breaking the symmetry of the out-of-plane direction and influencing both in-plane and out-of-plane phonon modes. The anomalous transition observed in germanene highlights its potential for tailored heat transport management, making it a promising candidate for thermoelectric applications where precise control over thermal properties is crucial \cite{sapta_germanene}.

A few works on the graphene/germanene (Gr/Ge) heterostructure are reported in the literature. The formation and stability of this vdW heterostructure are confirmed through \textit{ab-initio} density functional theory (DFT) calculations \cite{Cai2013, Hamid2021}. Application of this heterostructure in gas sensing is also explored \cite{Wu2023}. However, not much emphasis is given to understanding the interfacial thermal conductance in this heterostructure. Experimental works on the ITC are very challenging due to the nonequilibrium nature of the phonon distributions involved \cite{Liu2025}. As a result, most of the works on interfacial thermal conductivity are based on classical molecular dynamics (MD) simulations. One of the primary reasons for the few studies on the heterostructure may be due to a lack of accurate interlayer interaction potential, which is essential for the simulation of ITC. Although a lot of works have focused on developing the interlayer interactions for graphene \cite{Wen2018}, h-BN \cite{Ouyang2020}, and TMDs \cite{Naik2019}, no such potential exists for understanding the interactions between graphene and germanene.

To address this lack of suitable interlayer potential, we develop a pairwise potential using the data obtained from accurate DFT calculations. We use this potential to understand the ITC phenomena in the Gr/Ge heterostructure. Further, the tunability of the ITC with temperature and external strains is explored through phonon density of state calculations so as to understand the impact on ITC due to imperfect device implementations.

This work is organized as follows: Sec. \ref{sec:theory} provides an insight into the theoretical understanding and computational schemes. In Sec. \ref{sec:results}, we provide an in-depth description of the results obtained: Sec. \ref{sec:structure} presents the structural properties and formation of the Gr/Ge heterostructure. We discuss the development of interlayer potential for accounting for the vdW interactions between the layers in Sec. \ref{sec:potopt}, followed by ITC calculations and their tunability with external conditions in Sec. \ref{sec:itc}. Sec. \ref{sec:mechanisms} explores the mechanisms underlying the superior tunability of ITC, with a detailed analysis of the phonon density of state. We summarize the key understanding and its potential applications in \ref{sec:conclusions}.

\section{\label{sec:theory}Computational Details\protect }

\subsection{\label{sec:theory_opt}Optimization of Interlayer Potential}
We use \textit{ab-initio} DFT calculations to obtain the binding energy curve for the Gr/Ge heterostructures. We use the Perdew-Burke-Ernzerhof generalized gradient approximation (PBE-GGA) functional, as implemented in \textsc{Quantum Espresso}\cite{Giannozzi2007, Giannozzi2017} package for calculating the exchange-correlation energy. The ultrasoft pseudopotential \cite{uspp} from the solid state pseudopotential library \cite{sssp2018} is used for accommodating the interactions between the electrons and the core. We note that the projected augmented wave and ultrasoft pseudopotentials produce the same result for binding energy in our calculations, and as a result, the ultrasoft pseudopotential is used for optimum computational cost. The semi-empirical Grimme's DFT-D3 is used to address the weak vdW interactions between graphene and germanene layers \cite{dftd32010}. A plane wave cutoff of 50 Ry and a charge density cutoff of 500 Ry are found to produce sufficiently accurate binding energy (Table T1 in the supplementary material (SM)) and therefore are used for the calculations. Owing to the large unit cell, a Monkhorst-Pack \cite{mp1976} k-point grid of 2 $\times$ 2 $\times$ 1 is adequate to calculate the binding energy (Table T2 in SM). The systems are minimized until Hellmann-Feynman forces acting on the atoms are less than $10^{-4}$ eV.

The Large-scale Atomic/Molecular Massively Parallel Simulator (\textsc{LAMMPS}) is used for simulating systems using MD simulations. Under classical considerations, the Hamiltonian for the system is given by:
\begin{equation}
    H = T + \Phi^{\text{intra}}_{\text{Gr}} + \Phi^{\text{intra}}_{\text{Ge}}+\Phi^{\text{inter}}_{\text{Gr/Ge}}.
\end{equation}
Here, T represents the kinetic energy, $\Phi^{\text{intra}}_{\text{Gr}}$ and $\Phi^{\text{intra}}_{\text{Ge}}$ represent the intralayer interactions between the carbon atoms in the graphene layer (Gr) and germanium atoms in the germanene layer (Ge), respectively. The interlayer interactions between the two layers are represented by $\Phi^{\text{inter}}_{\text{Gr/Ge}}$.

We use the bond order Tersoff potential \cite{Tersoff1989} optimized by Lindsay et al. \cite{TersoffLindsay2010} for accounting for intralayer interactions between carbon atoms in the graphene layer. For the germanene layer, we use the Stillinger-Weber (SW) potential \cite{sw1985} parametrized by Jiang et al. \cite{Jiang2017, Thapliyal2024}. For the interlayer interactions between the germanene and graphene monolayers, we use the pairwise Lennard-Jones (LJ) potential given by:
\begin{equation} \label{eqn:lj}
    \Phi^{\text{inter}}_{\text{Gr/Ge}}(r_{ij})= 4\epsilon\left[\left(\frac{\sigma}{r_{ij}}\right)^{12}- \left(\frac{\sigma}{r_{ij}}\right)^{6}\right], r_{ij}<r_c.
\end{equation}
Here, $r_{ij}$ is the distance between $i^{\text{th}}$ carbon atom and $j^{\text{th}}$ germanium atom. $\sigma$ and $\epsilon$ represent the parameters of the potential. 

The binding energy between the layers is calculated as:
\begin{equation} \label{eqn:be}
E_b^{\text{DFT/MD}}= E_{\text{Gr/Ge}}^{\text{DFT/MD}} - E_{\text{Gr}}^{\text{DFT/MD}} - E_{\text{Ge}}^{\text{DFT/MD}}.
\end{equation}
Here, $E_b$ represents the binding energy between the layers, $E_{\text{Gr/Ge}}$ is the energy of the heterostructure, $E_{\text{Gr}}$ is the energy of the graphene layer, and $E_{\text{Ge}}$ is the energy of the germanene layer. The subscripts DFT and MD represent whether the energy is calculated from the \textit{ab-initio} calculations or the classical calculations, respectively. 

We use the conjugate gradient minimizer to optimize the following objective function:
\begin{equation}
    \chi^2 = \frac{1}{W}\int_0^\infty \left|E_b^{DFT} (r)- E_b^{MD}(r)\right|^2w(r)dr
\end{equation}
\begin{equation}
	\label{eqn:chisq}
	 \approx \frac{1}{W}\sum_{r_i}^{r_{cut}}|E^{\mathrm{DFT}}_\mathrm{b}(r_i) - E^{\mathrm{MD}}_\mathrm{b} (r_i)|^2 w(r_i),
\end{equation}
where $w(r)$ specifies the weight at an interlayer distance $r$ and W is the total weight. The choice of weight is not unique, and in our calculations, we use equal weightage to evaluate the objective function.

\subsection{\label{sec:theory_itc}Interfacial Thermal Conductance}
For the simulations with classical potentials, we minimize the systems with a damped dynamics method with a tolerance of $10^{-14}$ for energy and $10^{-14}$ eV/{\AA} for force \cite{fire}. The optimized systems are simulated in the isothermal-isobaric (NPT) ensemble for 3.5 ns so that the systems attain an equilibrium state. The thermodynamic quantities under the equilibration run are plotted in Fig. S1 in the SM.  It is noteworthy that the temperature of both the layers in the equilibration run must be monitored, as due to the unequal mass, the equilibration of separate layers will take much longer than the entire system, evident from Fig. S1(a) in the SM. After this equilibration run, the germanene layer is quickly heated to T+200 within 50 ps using the velocity rescale thermostat. The heated systems are then allowed to relax until a steady state is achieved in the NVE ensemble by removing the thermostats.

The interfacial thermal conductance between the layers is calculated as:
\begin{equation}
\label{eqn:itc}
    G=\frac{C_V}{A\tau},
\end{equation}
where, $C_V$ is the effective constant volume heat capacity, $A$ is the interfacial area, and $\tau$ is the thermal relaxation time.

The constant volume heat capacity is calculated using Phonopy \cite{phonopy} as: 
\begin{equation}
    C_V = \left(\frac{\partial E}{\partial T}\right)_V,
\end{equation}
where, $E$ is the total internal energy at temperature $T$ for the Gr/Ge system. In terms of phonon frequency, we can write:
\begin{equation}
    C_V = \sum_{\textbf{q}\nu}k_B\left(\frac{\hbar\omega(\textbf{q}\nu)}{k_BT}\right)^2\frac{\exp(\hbar\omega(\textbf{q}\nu)/k_BT)}{[\exp(\hbar\omega(\textbf{q}\nu)-1]^2},
\end{equation}
with $\omega$ is the angular frequency of phonon with a wavevector $\bm{q}$ in branch $\nu$, $k_B$ represents the Boltzmann constant, and $\hbar$ is the reduced Planck's constant.
The thermal relaxation time is calculated by fitting:
\begin{equation}
	\label{eqn:relax}
    \Delta T(t) = \Delta T(t_0)\exp\left[\frac{(t_0-t)}{\tau}\right],
\end{equation}
where, $t_0$ is the starting time of the thermal relaxation process, $\tau$ is the thermal relaxation time, and $\Delta T(t)$ is the temperature difference between the two sheets at time $t$.
We integrate the velocity autocorrelation function (VACF) to calculate the phonon density of state (PDOS) using the following:
\begin{equation}
	\label{eqn:dw}
	D(\omega) = \frac{1}{3Nk_BT}\int_{0}^{\infty}\frac{\langle \bm{v}(0).\bm{v}(t)\rangle}{\langle \bm{v}(0).\bm{v}(0)\rangle}e^{i\omega t}dt,
\end{equation}
where, VACF is given by: $\langle \bm{v}(0).\bm{v}(t)\rangle$ for phonons with angular frequency $\omega$, velocity $\bm{v}(t)$ at time $t$, and $N$ is the number of atoms. For these calculations of PDOS, we take a finer timestep of 0.05 fs and a data saving frequency of 0.2 fs for 250 ps in the NVT ensemble, which results in a smoother PDOS. 
\section{\label{sec:results}Results and Discussion}
\subsection{\label{sec:structure}Structural Geometry}
\begin{figure}[h]
	\includegraphics[height=6.5cm, keepaspectratio]{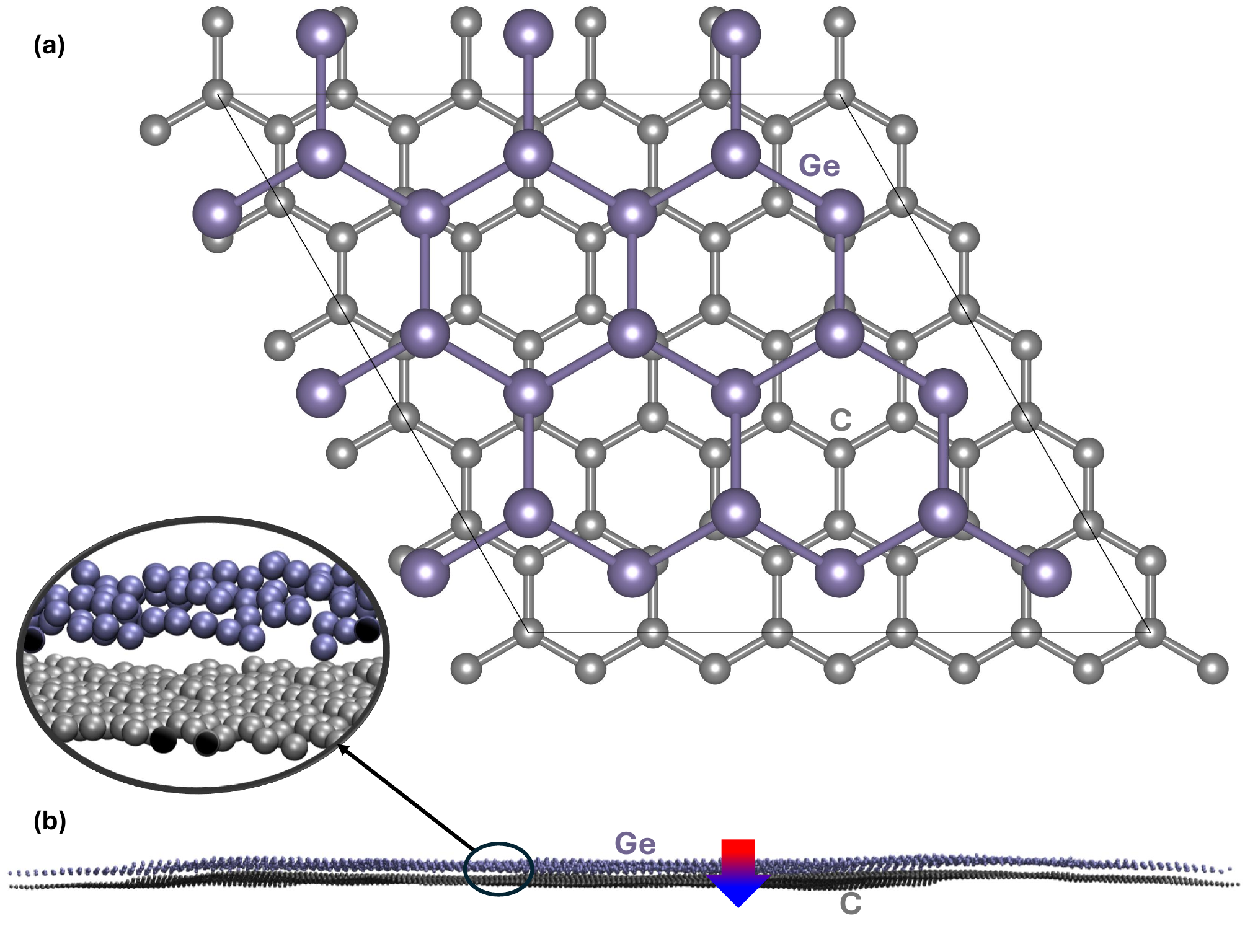}
	\caption{(a) The top view unitcell configuration of the Gr/Ge vdW heterostructure created by placing a 3$\times$3$\times$1 supercell of germanene monolayer on top of a 5$\times$5$\times$1 supercell of graphene monolayer and (b) the side view of the MD equilibrated 16 $\times$ 16 $\times$ 1 supercell of the heterostructure at room temperature. In the inset, we show a closer view of the same structure.}
	\label{fig:structure}
\end{figure}
For the construction of the Gr/Ge vdW heterostructure, we consider the optimized lattice structure for graphene and germanene monolayers from previous experimental and theoretical works \cite{CastroNeto2009, Bianco2013}. The graphene crystal has a planar structure with a lattice constant of 2.46 {\AA}, whereas germanene has a buckled structure with a lattice constant of 4.05 {\AA} and a buckling height of 0.71 {\AA}. Now, to construct a heterostructure from these two monolayers of mismatched lattice constant, we place a 3$\times$3$\times$1 supercell of germanene monolayer on top of a 5$\times$5$\times$1 supercell of graphene monolayer. The resulting structure contains a total of 68 atoms, with 18 Ge atoms and 50 C atoms, as shown in Fig. \ref{fig:structure}(a). The lattice constant of the constructed heterostructure system is 12.13 {\AA}, resulting in a strain of $\sim$1.5\% on both layers. The optimized interlayer distance between the two layers is found to be 3.52 {\AA} from vdW-corrected DFT calculations. The magnitude of interlayer separation is between the vdW diameters of C and Ge atoms and signifies the formation of a vdW heterostructure. For calculations of interfacial thermal conductivity, we extend this heterostructure to create a 16 $\times$ 16 $\times$ 1 supercell containing 17408 atoms; an equilibrated snapshot of the same at room temperature is shown in Fig. \ref{fig:structure}(b). This is consistent with previous works in graphene/silicene vdW heterostructure and is sufficient for eliminating finite size effects in classical MD simulations \cite{Liu2014}.

\subsection{\label{sec:potopt}Optimization of interlayer potential}
\begin{figure}[h]
	\includegraphics[height=6.5cm, keepaspectratio]{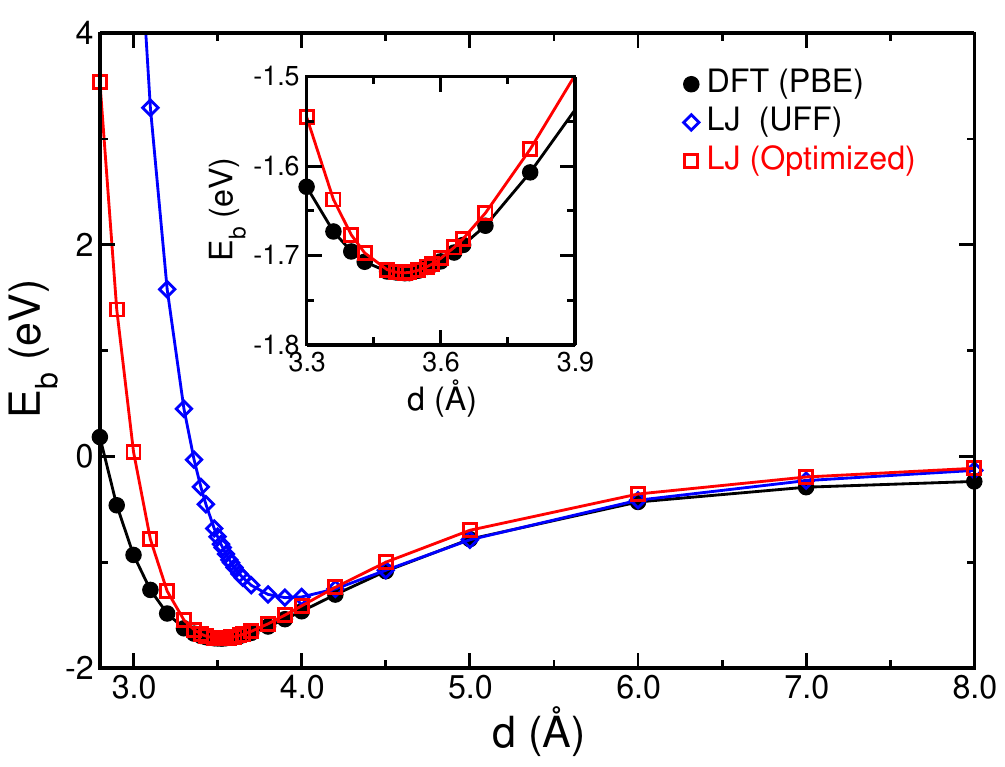}
	\caption{Binding energy for Gr/Ge heterostructure with interlayer distance calculated from DFT as well as classical molecular mechanics simulations. It is clear that the UFF model \cite{uff1992} cannot produce the minima of the binding energy as obtained from DFT calculations. We find the optimized LJ potential ($\epsilon$ = 0.017 eV and $\sigma$ = 3.67 {\AA}) accurately describes the minima of the heterostructure. In the inset, the data around the minima is shown for clarity.}
	\label{fig:be_curve}
\end{figure}
Understanding interfacial thermal conductance and similar phenomena involving phonon transport in vdW heterostructures is challenging due to the difficulty in accurately describing the interlayer interactions involved. Simulating a vdW heterostructure with a large lattice mismatch for phonon thermal transport is computationally infeasible using DFT due to the large number of atoms present in the heterostructure. For instance, the Gr/Ge vdW heterostructure in this work contains 68 atoms in the unit cell. As a result, it is imperative to design a suitable interlayer potential by using the data calculated from accurate DFT calculations. Here, we use the pairwise LJ potential to account for the interlayer interactions. It must be mentioned that, in general, a threebody potential like the Kolmogorov-Crespi (KC) \cite{Kolmogorov2005} or the dihedral-angle-corrected registry-dependent interlayer potential (DRIP) \cite{Wen2018} may be highly suitable due to their three-body nature. But, since the unitcell of the Gr/Ge vdW heterostructure is quite large so as to minimize the lattice mismatch between the layers, it contains all types of possible stacking for the two monolayers in the unitcell. As a result, the sliding energy is constant, and a threebody potential is not necessary. In Fig. \ref{fig:be_curve}, we plot the binding energy (calculated using Eq. \ref{eqn:be}) of the Gr/Ge heterostructure with interlayer distance calculated with both the DFT and classical (molecular mechanics) calculations.
The reference parameters for the fitting process (Eq. \ref{eqn:chisq}) are taken from Rappe's unified force field (UFF) \cite{uff1992}. A global cutoff of 20 {\AA} is used for the LJ potential, which is more than five times the $\sigma$.  It is found that in 2D materials, UFF does not provide an accurate description of the interactions as well as the interlayer distance due to inaccurate minima in the binding energy curve (\ref{fig:be_curve}), similar to other 2D materials \cite{sapta2023}. The conjugate gradient minimizer used to minimize the $\chi^2$ function obtains the LJ potential parameters to be $\epsilon$ = 0.017 eV and $\sigma$= 3.67 {\AA}. Our optimized parameters, along with the Tersoff potential for intralayer interactions of graphene and the SW potential for intralayer interactions of germanene, reproduce accurate interlayer distance (3.52 \AA) as well as buckling of germanene (0.71 \AA), as obtained from DFT calculations. As a result, for all further classical calculations, we use the optimized parameters to model the heterostructure. Here, we note that there is a slight deviation in the fitted binding energy curve using our optimized potential with the DFT data, beyond the minima. As a result, the calculations done for the strained structure may have some errors. However, the system with the largest compressive strain used in this work (5\%) will have a deviation of 0.08 eV, or 1.17 meV/atom in the binding energy, which might not have any significant effect on the result.

\subsection{\label{sec:itc}Interfacial Thermal Conductance and Its Tunability}
The interfacial thermal conductance is calculated from the relaxation of the rapidly heated system in the NVE ensemble, using Eq. \ref{eqn:itc}. More details on the calculations are explained in Sec. \ref{sec:theory_itc}. In Fig. S2, we present the temperature with time of both the Gr and Ge sheets after the thermostats are removed and the system is allowed to relax in the NVE ensemble. We find that the temperature of the Gr sheet sharply drops, and the Ge sheet rises. The Gr/Ge heterostructure system achieves a steady state at $\sim$370 K when the heat is transported from the Gr to the Ge sheet, i.e., during the heating process, the Gr sheet is heated to 500 K with the Ge sheet being kept at room temperature. When the heat flows in the other direction, i.e., when the Ge sheet is heated to 500 K and the Gr sheet is kept at 300 K, the average temperature of the system stabilizes at 320 K (Fig. S2(b)). The temperature difference at relaxation with simulation time in the NVE ensemble is plotted in Fig. S3 in the SM. The relaxation time is obtained by fitting this data to Eq. \ref{eqn:relax}, as shown in Fig. S3 in the SM.

\begin{figure}[t]
	\includegraphics[height=6.5cm, keepaspectratio]{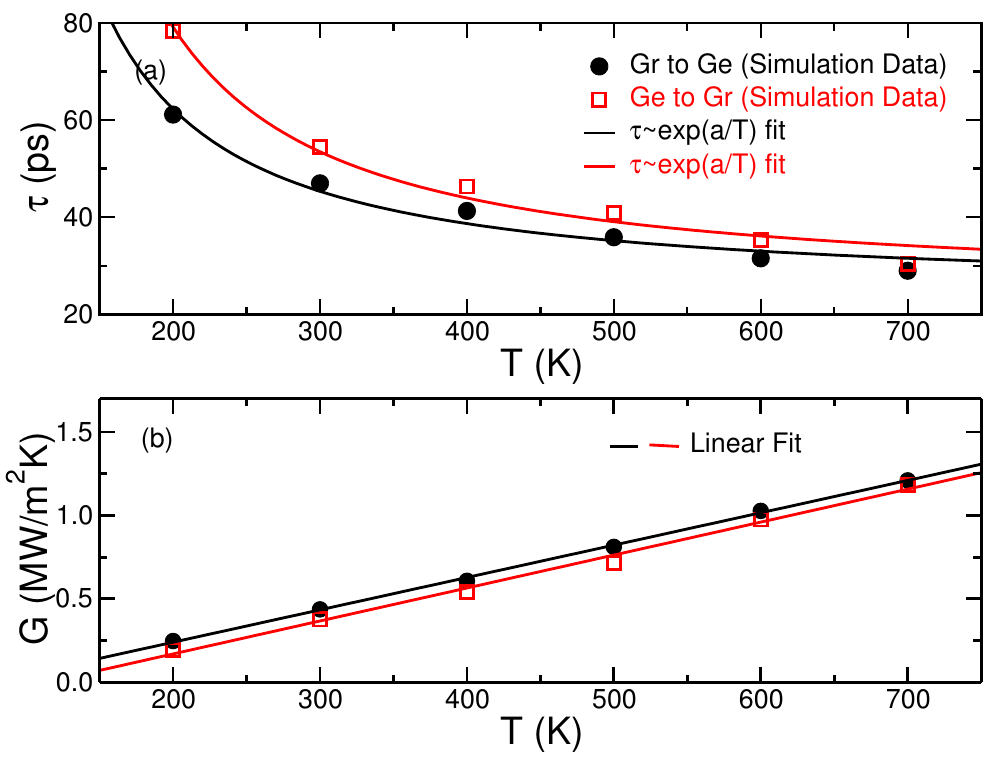} 
	\caption{(a) The thermal relaxation time with temperature for the heterostructure system. The black circles represent $\tau$, the thermal conduction from the graphene to the germanene layer, and the red squares represent the reverse conduction. It is observed that $\tau$ monotonically decreases with temperature and is consistently high in the case of conduction from the germanene to the graphene layer, and (b) interfacial thermal conductance of Gr/Ge heterostructure with temperature. We observe that the conductance increases monotonically with temperature, whether the heat is transferred from graphene to germanene or the other direction.}
	\label{fig:tau}
\end{figure}
We present the thermal relaxation time $\tau$ with temperature in Fig. \ref{fig:tau}(a). We find that the thermal relaxation time decreases monotonically with temperature. At 300 K, a thermal relaxation time of 47 ps is calculated when the heat is transferred from the graphene to the germanene sheet, while we find $\tau$ to be 54.5 ps in the opposite direction of heat conduction. This trend of higher relaxation time for germanene to graphene heat conduction holds for the whole range of studied temperatures, with the gap narrowing at higher temperatures. The temperature variation of the relaxation time may be fitted to an Arrhenius-like equation $\tau=\alpha\exp(\frac{\beta}{T})$.

In Fig. \ref{fig:tau}(b), we plot the interfacial thermal conductance with temperature. The interfacial thermal conductance is found to be 0.44 MW/m$^2$K for heat transfer from graphene to germanene and 0.38 MW/m$^2$K for heat conduction in the opposite direction at room temperature. This conductance is lower than that obtained for the graphene/silicene vdW heterostructure \cite{Liu2014}, likely due to the lower in-plane thermal conductivity of monolayer germanene resulting from its high buckled structure. As the temperature is raised, the thermal conductance is found to increase monotonically for both directions of heat transfer. The increase in ITC may be fitted with a linear equation, implying a steady increase with temperature. Similar to the case of relaxation time, the thermal conductance is found to be higher for Gr to Ge heat conduction. The increase in the ITC with temperature in both cases of heat transport is linear in nature and can be fitted to a linear equation as shown in Fig. \ref{fig:tau}(b).

\begin{figure}[b]
	\includegraphics[height=6.5cm, keepaspectratio]{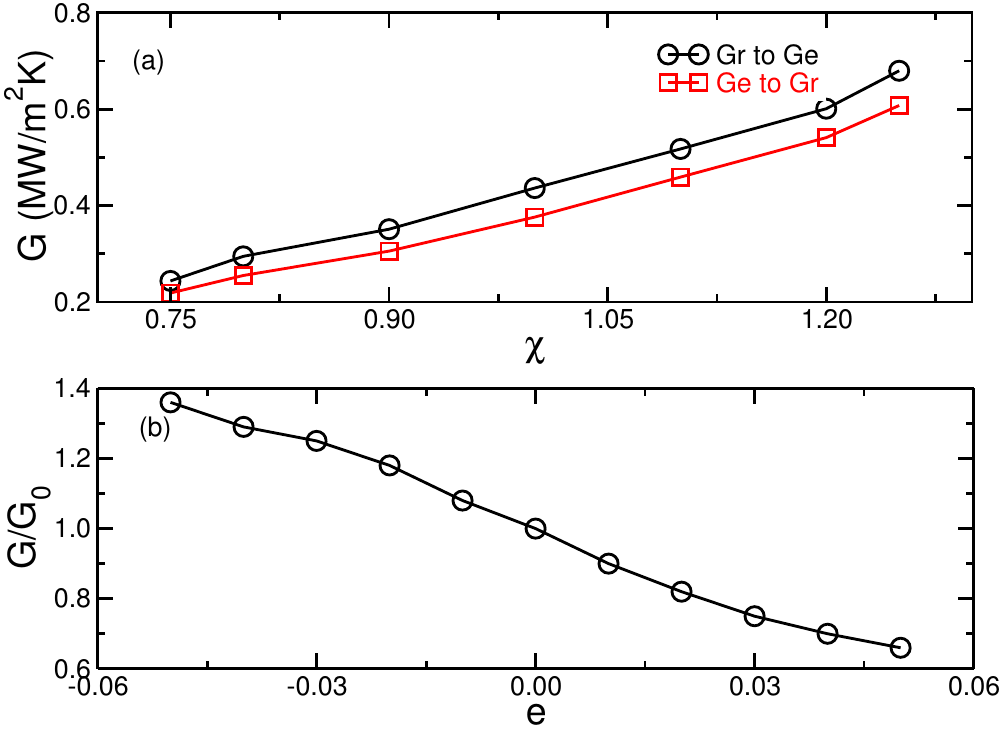}
	\caption{(a) Interfacial thermal conductance of Gr/Ge heterostructure with the interlayer interaction strength at room temperature. We find that the interfacial conductance increases monotonically with the interaction strength for both directions of heat transport. In (b), the variation of scaled interfacial thermal conductance with external strain at room temperature is shown for Gr to Ge heatflow. We note that a tensile strain decreases the interfacial thermal conductance, while a compressive strain enhances it.}
	\label{fig:chivsg}
\end{figure}
Modeling the realistic interactions in any vdW heterostructure is challenging due to the complexity of experimental realization. To account for the experimental variation in interactions as well as to understand the variation of interaction effect on the interfacial thermal conductance, we vary the LJ interaction (Eq. \ref{eqn:lj}) between the layers by using a dimensionless factor $\chi$ as:
\begin{equation}
    \Phi^{\text{inter}}_{\text{Gr/Ge}}(r_{ij})= 4\chi\epsilon\left[\left(\frac{\sigma}{r_{ij}}\right)^{12}- \left(\frac{\sigma}{r_{ij}}\right)^{6}\right], r_{ij}<r_c.
\end{equation}

We present the variation of ITC with the strength of interaction in Fig. \ref{fig:chivsg}(a) at room temperature for both heat conduction directions. We observe that interlayer interaction strength positively influences the ITC of the Gr/Ge heterostructure, similar to the case of the graphene/silicene heterostructure \cite{Liu2014}. When the interaction strength is lowered to 75\% of the optimized value, the ITC drops by $\sim$50\% at room temperature. On the other hand, when the interaction strength is raised to 125\% of the optimized value, the ITC is raised by 76\%.

In any implementation, devices are always susceptible to the effects of external strain. To understand the effect of strain on the conduction of heat current and thermal conductance, we apply both compressive and tensile strain in the heat flow direction. We define the cross-plane strain as:
\begin{equation}
    e = \frac{l-l_0}{l_0},
\end{equation}
where \textit{l} is the compressed/elongated length and \textit{l$_0$} is the equilibrium length in the direction of heat flow.

The variation of scaled (with respect to pristine structure) ITC due to external compressive as well as tensile strain is plotted in Fig. \ref{fig:chivsg}(b) for Gr to Ge heatflow at room temperature. We find that the ITC decreases from the pristine structure as a tensile strain is applied in the direction of heat flow. ITC follows a monotonic decrease as strain is increased. We find that the ITC is decreased to $\sim$70\% of that of the pristine structure when a 5\% tensile strain is applied at 300 K. Interestingly, ITC is found to increase when a compressive strain is applied in the direction of heatflow. The interfacial thermal conductance (ITC) increases to approximately 136\% of its value in the pristine structure upon the application of 5\% compressive strain.

\subsection{\label{sec:mechanisms}Mechanism Underlying Thermal Conductance}
\begin{figure}[h]
 \includegraphics[height=6.5cm, keepaspectratio]{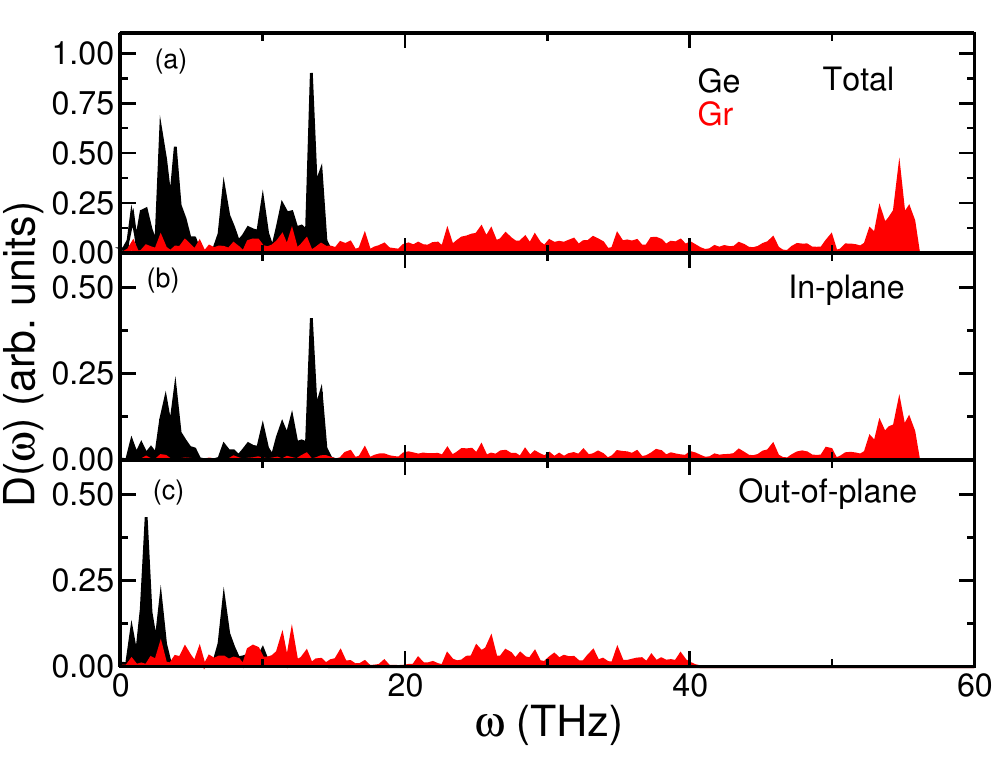}
 \caption{(a) The total phonon density of state with phonon angular frequency. The total PDOS is further decomposed into the (b) in-plane and (c) out-of-plane components. Here, the black curve represents the contribution from the Ge sheet, and the red curve represents the contribution from the Gr sheet. It is evident that the phonons overlap in the low-frequency region.}
 \label{fig:pdos}
\end{figure}
To understand the mechanisms of interfacial thermal conductance, we calculate the PDOS of the Gr/Ge heterostructure. We present the total PDOS of the heterostructure in Fig. \ref{fig:pdos}(a). It is observed that the low-frequency phonons of graphene overlap with the phonons of germanene. It is important to note that the out-of-plane phonons contribute significantly to the thermal transport in graphene. On the other hand, the out-of-plane phonons in germanene experience significant scattering due to the high buckling of the crystal structure. So, we decompose the PDOS into in-plane and out-of-plane contributions in Fig. \ref{fig:pdos}(b) and (c), respectively, to understand the contrasting behavior of in-plane and flexural phonons. It is quite clear that the out-of-plane component of Gr primarily dominates at the low-frequency regime and overlaps with the phonon modes of Ge. As a result, this coupling between low-frequency out-of-plane phonons of Gr and the phonons of Ge is the most significant mode of transport for heat conduction in this vdW heterostructure. 

When the temperature is low, fewer phonons are available in Gr for the transport of heat current \cite{Liu2014}. This leads to a lower coupling between the two layers, leading to a higher relaxation time. A higher relaxation time leads to a lower interfacial thermal conductance. As the temperature is raised, Umklapp processes are increased, and a higher number of phonons are available for the transport of heat current due to the scattering of excited high-frequency phonons to low frequencies. As a result, the thermal relaxation time is reduced, increasing the ITC of the heterostructure.

\begin{figure}[h]
	\includegraphics[height=6.5cm, keepaspectratio]{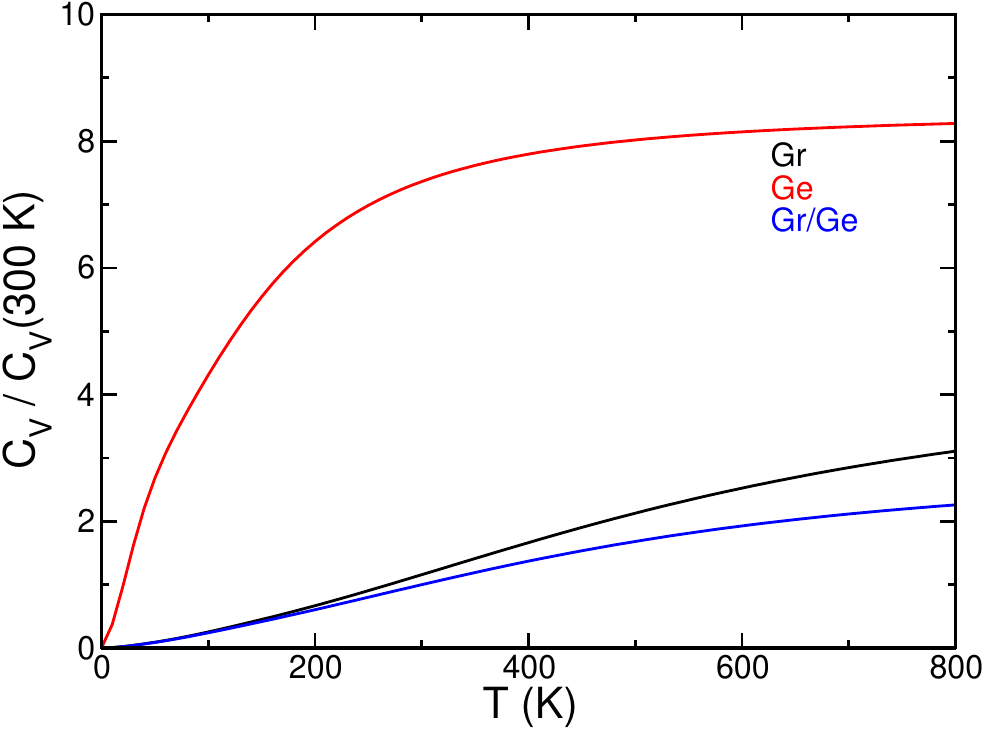}
	\caption{The relative constant volume specific heat capacity of monolayers as well as the Gr/Ge heterostructure with temperature. The specific heat capacities of all three systems (i.e., Gr monolayer, Ge monolayer, and the Gr/Ge heterostructure) are scaled with the specific heat capacity of the Gr/Ge heterostructure at 300 K.}
	\label{fig:cv}
\end{figure}

The anisotropic nature of ITC in the Gr/Ge heterostructure can be explained by the uneven heat capacity and atomic density of the layers. In Fig. \ref{fig:cv}, we plot the relative constant volume specific heat capacity of monolayers as well as the Gr/Ge heterostructure with temperature. The specific heat of all three systems (i.e., Gr monolayer, Ge monolayer, and the Gr/Ge heterostructure) is scaled with the specific heat of the Gr/Ge heterostructure at 300 K. As the temperature is raised, the heat capacities are also raised monotonically. It is clear that the rate of increase of relative specific heat is much higher in the graphene layer than in the germanene layer. For example, as the temperature is raised from 200 K to 400 K, we find the relative specific heat of graphene to increase by $\sim$150\%, compared to a $\sim$21\% increase in the case of germanene (Fig. \ref{fig:cv}). When the heat flow is from graphene to germanene, i.e., the graphene layer is heated to a higher temperature, there are many more phonons available for heat transport across the layer due to the higher heat capacity. The phonons of graphene are excited to a much higher frequency, and they scatter into low-frequency phonons, causing the increased heat transfer. On the other hand, in the case of germanene to graphene heat transport, fewer phonons are excited, causing less ITC compared to the other direction. Moreover, a higher density of atoms in the Gr layer due to a smaller lattice structure results in a higher number of phonon modes available in Gr with the increase in temperature compared to the Ge layer, resulting in the anisotropic thermal conductance. Further, when the interaction strength of the LJ potential is increased, the couplings between the phonons of both layers are increased, in addition to the increased couplings between the in-plane phonons and flexural phonons \cite{Liu2014}. This results in an increased thermal transport in the interface, leading to an increased ITC with the strength of interaction.
 
\begin{figure}[h]
	\includegraphics[height=6.5cm, keepaspectratio]{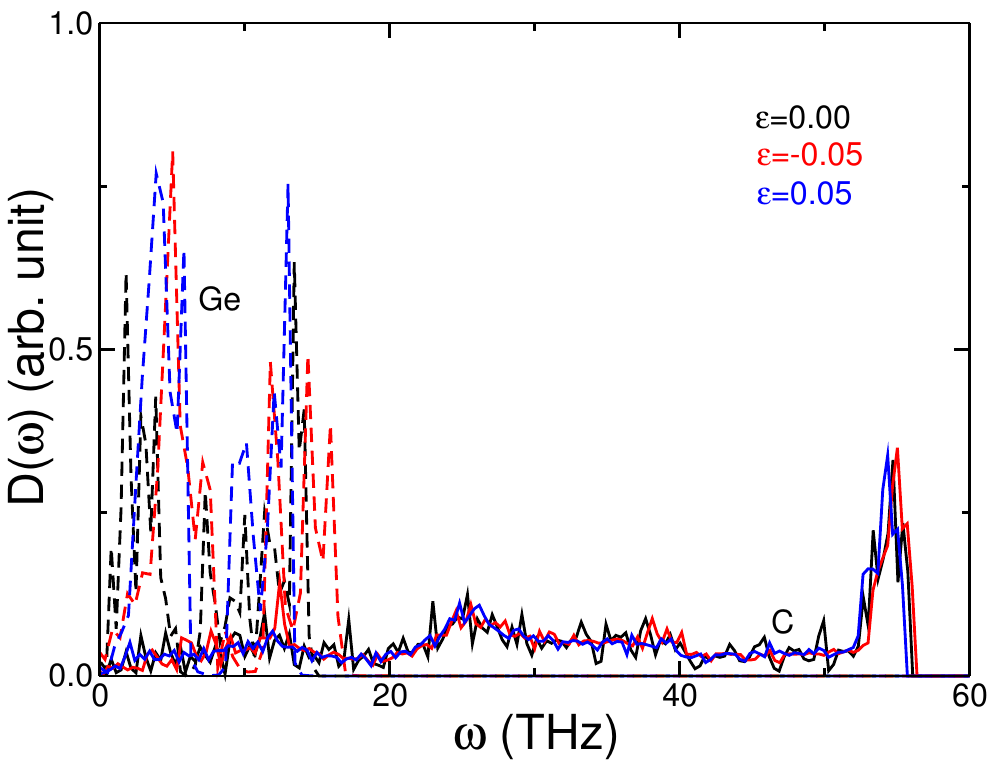}
	\caption{The total phonon density of state with phonon angular frequency for different external strains. The colored region shows the phonon contribution due to the germanene monolayer, while the lines represent graphene. It is evident that the overlap between the two phonon spectra increases with compressive strain, while it is reduced when a tensile strain is applied.}
	\label{fig:pdos_strain}
\end{figure}
In Fig. \ref{fig:pdos_strain}, we present the total PDOS of the Gr/Ge heterostructure for pristine and 5\% compressive and tensile strained structures at 300 K. Moreover, to provide a simplistic quantification of the overlap, we calculate the relative maximum phonon frequency percentage as $(\omega_{Ge}^{max}/\omega_{Gr}^{max})\times 100$ in Fig. S4 in the SM. We find that a compressive strain applied in the heat flow direction causes a blueshift of phonons of both graphene and germanene. As a result, the overlap region between the phonon modes of germanene and the low-frequency phonons of graphene increases substantially. Therefore, more channels are available for heat transport between the two layers. This is the reason for the monotonic increase in the ITC with an increase in the compressive strain. On the other hand, a tensile strain causes the phonons to redshift, leading to a decreased overlap between the phonon modes of germanene and the low-frequency phonons of graphene. Tensile strain causes the interlayer interaction to decrease, which directly correlates to a decoupling between the phonons of graphene and germanene. Therefore, a reduction in heat transport between the layers is observed.

\section{\label{sec:conclusions}Conclusions\protect}
In summary, we develop an accurate pairwise interlayer interaction potential for the Gr/Ge vdW heterostructure using binding energy data obtained from the first-principles calculations. When used in classical simulations, the optimized potential accurately describes the lattice parameters as well as the interlayer separation between the two layers in the vdW heterostructure. The pump-probe-based simulations calculate the ITC to increase significantly with compressive strain in the direction of heat flow, due to an overall increase in phonon modes available for heat transport. On the other hand, decreased interlayer interaction leads to a redshift in phonon modes, causing an overall decrease in the ITC with the increase in tensile strain. An increase in temperature or interlayer coupling strength enhances the ITC in the heterostructure, primarily due to the greater availability of phonon modes that facilitate heat transport. The ITC exhibits anisotropic behavior, which stems from the intrinsic anisotropy in the constant-volume heat capacity and atomic density of the constituent monolayers. 
The optimized potential model developed in this work will be helpful in accurately modeling the physical properties, such as relaxation and the phonon dispersion of the heterostructure and its twisted layers. Further, the findings reported here will provide great insights into designing thermal interfacing devices based on this heterostructure.
\section*{SUPPLEMENTARY MATERIAL}
The supplementary material contains (S1) Optimization of DFT parameters, (S2) equilibration of the Gr/Ge heterostructure in the NPT ensemble, (S3) relaxation of the Gr/Ge heterostructure in the NVE ensemble, (S4) exponential fit to relaxation of the Gr/Ge heterostructure, and (S5) the relative maximum phonon frequency percentage with external strain.
\begin{acknowledgments}
	The authors acknowledge the Computer Center of IIT Jodhpur for providing computing resources that contributed to the research results reported in this paper.
\end{acknowledgments}
\section*{CRediT author statement}

\textbf{Sapta Sindhu Paul Chowdhury}: Conceptualization, Methodology, Software, Investigation, Formal analysis,  Writing - Original draft preparation, Visualization; \textbf{Sourav Thapliyal}: Investigation, Validation, Writing - Review \& Editing, Visualization; \textbf{Bheema Lingam Chittari}: Methodology, Writing - Review \& Editing; \textbf{Santosh Mogurampelly}: Conceptualization, Methodology, Formal analysis, Writing - Review \& Editing, Supervision.

\section*{Conflicts of interest}
The authors have no conflicts to disclose.

\section*{Data Availability Statement}
The data that supports the findings of this study are available within the article and its supplementary material.
\bibliography{gr_ge_itc}
\end{document}